**Contemplating Secure and Optimal Design Practices for Information Infrastructure From a Human Factors Perspective**

Niroop Sugunaraj

School of Electrical Engineering and Computer Science (SEECS)

College of Engineering and Mines (CEM)

University of North Dakota

March 2024

**Abstract**

Designing secure information infrastructure is a function of design and usability. However, security is seldom given priority when systems are being developed. Secure design practices should balance between functionality (i.e., proper design) and usability to meet minimum requirements and user-friendliness. Design recommendations such as those with a user-centric approach (i.e., inclusive of only relevant information, user liberty) and presenting information within its proper context in a clear and engaging manner has been scientifically shown to improve user response and experience.

**Contemplating Secure and Optimal Design Practices for Information Infrastructure From a Human Factors Perspective**

**Introduction**

Human-engineered systems are primarily designed to follow a three-pronged approach: a system that focuses on users, built with a purpose, and/or resilient enough to avoid failures (Remington et al., 2012). Firstly, designing user-centric systems considers multiple factors such as learnability, efficiency, memorability, objective measures of evaluation (e.g., error frequency), and subject measures of evaluation (e.g., user satisfaction) (International Standards Organization, 1998). Secondly, systems built with a purpose have a clear-cut reason for why they're engineered for use. For instance, hearing aids are designed to amplify ambient sounds for those with impaired auditory perception; since audition is one of four inputs to sensory memory (the other three being gustatory, iconic, and haptic), it is a critical piece that is needed for decision-making in the natural world. Lastly, systems are designed with the intention to avoid failure because designing such systems takes resources (i.e., human and financial) so failures in system operation can lead to significant financial burdens or even loss of life (e.g., motor vehicles).

However, not all systems or components within such systems rigorously meet all the criteria in this three-pronged approach and there is ample evidence of the contrary, particularly when considering security. Findings from a study (Furnell et al., 2007) reports that users who, despite being aware of the consequences of poor security, cite constraints such as security is too expensive, impedes the use of a system (i.e., computer in this case), and not having enough time to implement security practices; these reasons implictly indicate that security is not adequately prioritized and therefore requires some intervention from the developer-side to partially make it feasible for users to make better security choices.

As highlighted by Pfleeger and Caputo (Pfleeger and Caputo, 2012), a realistic cyber security scenario is one where architects or developers of system architectures should be cognizant of behavioral sciences as they design, develop, and use them to enable users to be at the center of implementing good security practices and not at the periphery. For example, the relevance of behavioral sciences to security can be factored through aspects such as biases (e.g., confirmation biases that lead to noticing and implementing evidence that supports a previously-held opinion) and decision-making (e.g., taking actions that underestimate risks that are willingly taken versus those that are out of an agent's control) (Schneier, 2008) which are aspects important to security.

     This report will critically evaluate the claims made by in the case study presented by Pfleeger and Caputo and offer recommendations for developers to collectively enhance the security of information infrastructure.

## Development

     The case study highlights a critical aspect of modern information infrastructure and that is the often secondary consideration of security in system development and usage. It highlights how users and developers primarily focus on essential but typically over-emphasized tasks such as information retrieval, analysis, and productivity and therefore relegate security as a peripheral concern. Moreover, security mechanisms are typically chosen based on technical feasibility and system objectives, neglecting user usability and cognitive load. This can lead to security mechanisms that are ineffective, inefficient, or even counterproductive, as they may interfere with the user's primary tasks, increase their cognitive load, or reduce their trust in the system. Recognizing this gap, it's imperative to enhance security awareness among system builders to ensure the effectiveness and utility of security measures.

Design Recommendations

     While this is a challenging issue that requires a careful balance between security and usability, the recommendations to the engineers who are designing the system will be constrained to the following:

1. User-centric approach: Involve the end users in the design process from the beginning. Understanding their needs and expectations for the system, giving them freedom (i.e., choice) to implement security, and gathering feedback and insights are practical methods to the user-centric approach. Also, user-centered design principles and best practices such as establishes guidelines and standards for user interface design (e.g., consistency, visibility) to create a system that is dependable will contribute to achieving sustainable long-term security goals.

2. Alerting systems: Design effective alerts and alarms that inform the user of any security issues or incidents that may affect their work or data. Use appropriate modalities (e.g., visual, aural, or tactile cues) to attract the user's attention and convey urgency. This also entails using meaningful symbols, colors, sounds, and words to communicate the nature and severity of a security mechanism.

**Presentation Recommendations**

Presenting the necessary information for cyber hygiene is as essential as designing defense-in-depth security measures that are typically performed by system architects. The following are presentation recommendations to optimize user perception and implementation:

1. Clear communication: Present security information in a clear and easily understandable manner, avoiding technical jargon or complex terminology. This also means using intuitive visual aids and user-friendly language to convey security alerts and instructions.

2. Contextual relevance: Tailor security alerts and alarms to the specific context of user activities. Provide relevant information about potential threats or security risks, along with actionable steps that users can take to mitigate them.

3. User engagement: Engage users to make informed decisions about security by providing them with not just the necessary information and resources, but also that maintains user attention out of their interest.

**Analysis**

This section will critically analyse each recommendation under the design and presentation categories for their feasibility and jot down any limitations that may exist. The interactions between design and presentation recommendations are quite obvious but best efforts will be made to address each recommendation as independently as possible.

Design Recommendations

In addition to understanding the needs and expectations of end-users, deploying an optimal user-centric design approach has to have some degree of autonomy that is given to users. As stated by Hof (Hof, 2015), security mechanisms should allow users to add, modify, or remove certain mechanisms that they may not deem necessary. More specifically, users should be given some degree of freedom to choose among a set of developer- or architect-approved security practices (e.g., approved web browsers or security virtual private networks (VPNs)) to use. Rather than allowing the user to subscribe to policies that follow black-box principles that the user may not understand, giving users the choice and (if required) the know-how behind security options/choices, users will be less focused on making good security decisions based on necessity rather than volition; this is a concept called information-flow security (Vachharajani et al., 2004) that gives users permission to audit or control programs within a system after giving them access.

Additionally, architects or developers should minimize the amount of information that is required by users to implement good security practices. For example, the reason why an average user uses the same password on multiple websites or accounts is because it reduces the memory footprint and hence the retrieval necessary to, say, login successfully. Using technologies that integrate seamless authentication without the need for passwor storage through services such as OpenID makes it easier and secure (balancing security and usability) for the system and its users (Hof, 2015).

By stating what constitutes good alerts and alarms, it is important to first differentiate alerts and alarms. Alerts refer to messages that inform a user or developer that divert his/her attention to a developing situation (i.e., less urgent) whereas alarms are messages that warrant immediate action (i.e., more urgent) from its agent. An example of an alert might the software on a personal device (e.g., smartphone) is out-of-date and a new version can be installed. An example of an alarm could be that a user's email was breached and a number of malicious emails were sent from this email.

Depending on the context and the physical environment, consideration should be given to whether the alert or alarm system's response modality is visual or aural (Remington et al., 2012). From a practical perspective, the ear is more sensitive to specific frequencies than others given the same amplitude; these frequencies are in the range of 1000 - 4000 Hertz (Hz) (Wilcox, 2011). Additionally, frequencies that are too close to one another are more difficult to hear. These two findings suggest that systems designed for alerts or alarms that have frequencies reasonably spread apart at the mid-range (i.e., 1000 - 4000 Hz) and ideally for different applications (e.g., automatic computer updates, phishing email notification).

Research done for multiple alarm modalities (i.e., auditory and visual) by Bliss and colleagues (Bliss et al., 1995) indicates that in cases where systems issue false alarms quite frequently (perhaps due to poor configurations/policies or overly sensitive systems), the cry-wolf effect (i.e., tendency to react to false positives and not true positives) leads to a degradation of response in terms of increased physiological stress and predicted response behavior. Therefore, it goes without saying that developers should be more acute and deliberate when designing stimuli or criteria that triggers such alarms, and refrain from being overly strict for the triggers.

**Presentation Recommendations**

Kelman's framework is relevant to this discussion for good presentation, because the research states that changes in behavior are motivated by a 3-part framework of compliance, identification, and internalisation. Compliance is motivated by shallow factors such as rewards/punishments for a particular behavior change and internalisation is the highest form of behavior change where changes in behavior are influenced by aligning the individual with their own personal values. Presenting information (e.g., security messages) that is engaging and makes implementation from the user side more feasible (i.e., path of least resistance) are key ways to harness the first (i.e., clear communication) and third (i.e., empowerment) recommendations. More practically, research demonstrates that having a video-style mini-series that presents security training in a popular streaming service format (e.g., Hulu, Netflix) or having a good stories included in security training that encourages skepticism, critical thinking, and creates emotional responses (i.e., with real-life situations) makes it more likely for users to not just engage but also retain the information presented (Alshaikh and Adamson, 2021).

Additionally, presenting security information (i.e., in a learning context) that is relevant and not redundant ensures that users are not fatigued or cognitively overloaded. As mentioned by Nobles (Nobles, 2018), majority of the cyber incidents in the U.S. and the U.K. are due to human errors; this can be significantly reduced by creating an organizational culture that implements security practices and processes that are focused on users in terms of increasing human performance and promote better decision-making. Supporting this fact is research that points to unintentional violations (e.g., shortcuts) primarily being due to users being necessitated to complete a certain task (Lawton, 1998) rather than them doing it of their own volition. Pressures of time, information overload leading to higher response times (i.e., Hick-Hyman law), and finding shortcuts or quicker ways of completing security training have all indicated higher likelihoods of users engaging in risky behaviors. Therefore, limiting information to include only the "necessary" aspects frees up user time and efforts and places a lower cognitive burden on them.

Presenting information optimally can be done in a variety of different ways but what will be briefly discussed is the proximity compatibility principle (PCP) first stated by Wickens and Carswell (Wickens and Carswell, 1995). PCP can be explained through its components of perceptual proximity (i.e., proximity in space for two displays providing task-related information) and processing proximity (i.e., the proximity that quantifies how two or more sources are used as part of the same task). Research lends support to the fact that perceptual proximity is disrupted when there is added clutter (e.g., irrelevant displays during a task); this means that visual scanning is disrupted and there is an added burden for users of scanning over and filtering out irrelevant information. Also, processing proximity offers benefits in cases where multiple characteristics (e.g., color, location on the screen) are integrated for an object on the display/screen (e.g., security training pop-up). These benefits (in terms of user response and experience) are evident when each there are multiple characteristics that each carry a certain implicit message (e.g., color of red with a sense of urgency, locating relevant information in the middle of the screen).

<p style="text-align:center"><strong>Conclusion</strong></p>

Systems designed by humans cater to specific requirements for which they are built. This indicates that such systems are purpose-driven, have a specific user-base, and to some degree are built to be resilient (i.e., failure-free). However, security is typically not considered during the design phase and lacks forethought. Therefore, secure design practices can be implemented by developers at the design and presentation levels to better cater to the system and its users. Design recommendations should be user-centered and focus on working through the needs of the user-base through presenting only relevant information, transparency, and liberty for users to make their own decisions within certain established bounds. Presentation recommendations should focus on communicating information clearly, be engaging, and relevant to the task at hand without cognitively overloading users with redundant information. Implementing alerts and alarms (especially in multiple modalities) should be given careful consideration and not be too sensitive (i.e., higher false positives); additionally, designing alerts/alarms to be spread apart over the mid-range frequency range caters better to the human aural function and elicits better responses from users.

## References

Alshaikh, M., & Adamson, B. (2021). From awareness to influence: Toward a model for improving employees' security behaviour. *Personal and Ubiquitous Computing*, *25*(5), 829–841.

Bliss, J. P., Gilson, R. D., & Deaton, J. E. (1995). Human probability matching behaviour in response to alarms of varying reliability. *Ergonomics*, *38*(11), 2300–2312.

Furnell, S. M., Bryant, P., & Phippen, A. D. (2007). Assessing the security perceptions of personal internet users. *Computers & Security*, *26*(5), 410–417.

Hof, H.-J. (2015). User-centric it security-how to design usable security mechanisms. *arXiv preprint arXiv:1506.07167*.

International Standards Organization, . (1998). Ergonomic requirements for office work with visual display terminals (vdts). https://www.iso.org/obp/ui/en/#iso:std:iso:9241:-1:ed-2:v1:en

Lawton, R. (1998). Not working to rule: Understanding procedural violations at work. *Safety science*, *28*(2), 77–95.

Nobles, C. (2018). Botching human factors in cybersecurity in business organizations. *HOLISTICA–Journal of Business and Public Administration*, *9*(3), 71–88.

Pfleeger, S. L., & Caputo, D. D. (2012). Leveraging behavioral science to mitigate cyber security risk. *Computers & Security*, *31*(4), 597–611. https://doi.org/https://doi.org/10.1016/j.cose.2011.12.010

Remington, R., Folk, C. L., & Boehm-Davis, D. A. (2012). *Introduction to humans in engineered systems*. John Wiley & Sons.

Schneier, B. (2008). The psychology of security. *International conference on cryptology in Africa*, 50–79.

Vachharajani, N., Bridges, M. J., Chang, J., Rangan, R., Ottoni, G., Blome, J. A., Reis, G. A., Vachharajani, M., & August, D. I. (2004). Rifle: An architectural framework for user-centric information-flow security. *37th International Symposium on Microarchitecture (MICRO-37'04)*, 243–254.

Wickens, C. D., & Carswell, C. M. (1995). The proximity compatibility principle: Its psychological foundation and relevance to display design. *Human factors*, *37*(3), 473–494.

Wilcox, S. B. (2011). A human factors perspective: Auditory alarm signals. *Biomedical Instrumentation & Technology*, *45*(4), 284–289.